\documentstyle[aps,prl,multicol,epsfig,amssymb,overcite]{revtex}

\begin{document}
\title{Signatures of a Concentration Dependent Flory $\chi$ Parameter:
Swelling and Collapse of Coils and Brushes}
\author{V.A. Baulin and A. Halperin\footnote{To whom
correspondence should be addressed, e-mail:
halperin@drfmc.ceng.cea.fr}}
\address{UMR 5819 (CEA, CNRS, UJF),
DRFMC/SI3M, CEA-Grenoble, 17 rue des
Martyrs,\\
38054 Grenoble Cedex 9, France}
\maketitle
\widetext

\begin{abstract}
\begin{center}
  \parbox{14cm}{
The quality of solvents of polymers is often described in terms of
the Flory $\chi $ parameter typically assumed to depend only on
the temperature, $T$. In certain polymer-solvent systems fitting
the experimental data enforces the replacement of $\chi(T)$ by a
concentration dependent $\chi_{eff}$. In turn, this modifies the
swelling and collapse behavior. These effects are studied, in the
framework of a mean-field theory, for isolated coils and for
planar brushes. The $\phi $ dependence of $\chi _{eff}$ gives rise
to three main consequences: (i) Shift in the cross-over between
Gaussian and self-avoidance regimes; (ii) A possibility of
first-order collapse transition for isolated flexible coils; (iii)
The possibility of a first-order phase transition leading to a
vertical phase separation within the brush. The discussion relates
these effects directly to thermodynamic measurements and does not
involve a specific microscopic model. The implementation for the
case of Poly(N-isopropylacrylamide) (PNIPAM) is discussed.}
\end{center}
\end{abstract}

\begin{multicols}{2}

\section{Introduction}

The quality of solvents of polymers is often quantified in terms
of the Flory free energy. In this free energy, the mixing energy
term has the form $ \chi \phi (1-\phi)$ where $\chi$ is the Flory
interaction parameter and $ \phi $ is the monomer volume fraction.
$\chi$ and the related second virial coefficient, $v=1-2\chi $,
measure the solvent quality. Thus $v>0$ corresponds to a good
solvent, $v=0$ to a $\theta $ solvent while $v<0$ indicates a poor
one. An important characteristic of these three regimes is the
associated swelling behavior. Thus, the span of an isolated chain
in the three regimes scales respectively as $N^{3/5}$, $N^{1/2}$
and $N^{1/3}$. Typically, it is assumed that $\chi $ depends only
on the temperature, $\chi =\chi (T)$.\cite{Flory,PGG,GK,Doi}
However when thermodynamic data are analyzed in terms of the Flory
free energy, it is often necessary to replace $\chi \phi (1-\phi
)$ by $\chi _{eff}\phi (1-\phi)$ where $\chi_{eff}$ is a function
of both $T$ and $\phi $ {\it i.e.}, $\chi_{eff}=\chi
_{eff}(T,\phi)$.\cite{Flory,wolf,Molyneux,Huggins,nom} The
introduction of $\chi_{eff}(T,\phi)$ requires, in turn,
modification of the discussion of solvent quality. In the
following we aim to clarify this issue by exploring some of the
microscopic consequences of $\chi_{eff}(T,\phi)$.

The solvent quality of polymer solutions as characterized by $\chi
_{eff}(T,\phi)$ concerns two issues: First, is the stability of
the solution with respect to phase separation due to diffusion of
chains. The second is the degree of swelling of polymer coils.
While the first issue was analyzed in considerable
detail,\cite{n-Cluster,Qian,Solc1,Solc2} less attention was given
to the second topic.\cite{BBP1,JK} The purpose of this article is
to explore the signatures of $\chi _{eff}=\chi _{eff}(T,\phi)$ on
the swelling behavior of isolated coils and of brushes of
terminally anchored chains. In turn, these signatures of the
solvent quality involve two aspects. One is a {\em ``global''}
solvent quality as revealed in systems of {\em uniform} density.
To study this facet we consider the effect of $\chi _{eff}(\phi)$
on the swelling behavior in two cases: (i) an isolated coil within
the Flory approximation\cite{Flory,PGG} and (ii) a brush as
described by the Alexander model.\cite{SA,HTL} As we shall
discuss, the global solvent quality is characterized by
$\overline{\chi }(\phi)=\chi _{eff}-(1-\phi)\partial \chi
_{eff}/\partial \phi $ rather than by $\chi _{eff}(\phi)$. The
$\phi $ dependence of $\overline{\chi }$ gives rise to two
effects. First is a qualitative modification of the collapse
transition that can assume, within these models, the character of
a first-order phase transition. Second is a significant shift in
the cross-over between the behavior of an ideal chain to that of a
self-avoiding one. For systems involving {\em gradients} in $\phi
$, it is necessary to consider a {\em ``local''} solvent quality.
In such systems $\overline{\chi }(\phi)$ is no longer the sole
measure of the solvent quality. The local solvent quality is
explored by using the Pincus approximation for
brushes\cite{Pincus,Pincus1} thus allowing for the concentration
profile and its coupling with $\chi _{eff}(\phi)$.\cite{footX} The
$\phi $ dependence of $\chi _{eff}$ gives rise to deviations from
the parabolic $\phi $ profile as obtained when $\chi =\chi (T)$.
In certain cases it leads to a first-order, vertical phase
separation within the brush. While this scenario was already
studied for the special case of polymers described by the ``{\em
n}-cluster'' model,\cite{Wagner,AHcluster} it is in fact a general
feature of systems characterized by $\overline{\chi }(\phi)$ that
increases with $\phi $. In all the cases listed above we compare
the scenarios found with $\chi _{eff}(\phi)$ to those resulting
from $\chi =\chi (T)$. Our discussion is based on simple
mean-field models and scaling refinements are ignored. This simple
minded approach is justified as an early step in analyzing the
problem. Clearly, a more sophisticated analysis may lead to
modifications of the results, especially with regard to the
collapse transition.

In studying the signatures of $\chi_{eff}(\phi)$ one may adopt two
strategies. One is to consider the problem within a specific
microscopic model.\cite{n-Cluster,Freed,Painter,Polysoaps,BH}
While this approach allows to trace the physical origins of the
effects, it suffers from two disadvantages. First, such analysis
is limited to the $\chi _{eff}(\phi)$ predicted by the particular
model and is only relevant to systems where this model is
physically reasonable. For example, some mechanisms are applicable
to all polymeric solutions\cite{Freed,Painter} while others
operate only for solutions of associating polymers\cite{Polysoaps}
or of neutral water-soluble polymers.\cite{BH} Second, each of the
microscopic models proposed thus far introduces extra parameters
that are presently unknown thus making confrontation with
experiments difficult. The second approach involves a
phenomenological description utilizing $\chi _{eff}(\phi)$ as
obtained from the colligative properties of polymer solutions.
This strategy limits the physical understanding of the swelling
behavior but allows to relate it to the solutions' colligative
properties as observed experimentally. In the following we pursue
the second, phenomenological approach. At present, the number of
polymer-solvent system for which $\chi _{eff}(\phi)$ data is
available is rather small. With this in mind we explore three
routes: (i) investigate the consequences of certain experimentally
measured $\chi _{eff}(\phi)$ curves (ii) utilize
$\chi_{eff}(T,\phi)$ obtained from empirical equations whose
parameters are determined by fitting the calculated phase diagram
with the observed one (iii) study the signatures of hypothetical
$\chi_{eff}(\phi)$ curves leading to qualitatively novel behavior.

Background material on the thermodynamics of polymer solutions
with $\chi _{eff}(\phi)$ is provided in section II. This section
also includes a unified description of the Flory approximation for
coils and the Alexander model for brushes. The next two sections
present an analysis of the swelling and collapse behavior within
these uniform density models. Thus, section III describes the
effects of $\chi _{eff}(\phi)$ on the swelling behavior of
isolated coils utilizing the Flory approximation while section IV
presents a similar discussion for the case of a planar brush as
described by the Alexander model. The coupling of $\chi
_{eff}(\phi)$ with a spatially varying $\phi$ is discussed in
section V, using the Pincus approximation. In section VI the
results of sections III-V are implemented for the case of
Poly(N-isopropylacrylamide)(PNIPAM) using an empirical expression
for $\chi _{eff}(T,\phi)$. In the final section we reconsider the
relative merits of the phenomenological and microscopic approaches
to the investigation of $\chi _{eff}(T,\phi)$ and its signatures.

\section{Coils and Brushes -- The Flory and Alexander Approximations with $
\protect\chi _{eff}(\protect\phi)$: Background}

In this section we first summarize the thermodynamics of polymer
solutions characterized by $\chi _{eff}(T,\phi)$. We than present
a brief unified description of the Flory and Alexander
approximations. In this, we focus on the balance of osmotic
pressure and elastic restoring force in determining the swelling
behavior. This presentation makes for a direct relationship
between the macroscopic thermodynamic properties and the
microscopic swelling behavior.

The replacement of $\chi =\chi (T)$ by $\chi_{eff}=\chi
_{eff}(T,\phi)$ requires certain modifications in the
thermodynamics of polymer solutions\cite{wolf} utilizing a
Flory-like mixing free energy. The mixing free energy per lattice
site, $f$, consists of two terms $f=f_{int}+f_{trans}$. One is an
interaction free energy $f_{int}/kT=\chi _{eff}(\phi)\phi (1-\phi
)$ that is the counterpart of the mixing energy $\chi \phi (1-\phi
)$. The second is the familiar translational free energy
$f_{trans}/kT=\phi /N\ln \phi +(1-\phi)\ln (1-\phi)$. As usual,
the chemical potential of the solvent is $\mu _{s}=\mu
_{s}^{o}(P,T)-\pi a^{3}$, where $\pi =a^{-3}\phi ^{2}\partial
(f/\phi)/\partial \phi $ is the osmotic pressure
\begin{equation}
\frac{\pi a^{3}}{kT}=\frac{\phi }{N}-\phi -\ln (1-\phi
)-\overline{\chi } \phi ^{2} \label{Piosm}
\end{equation}
and $a^{3}$ is the volume of a unit cell in of the lattice.
However, $\pi $ now depends on $\overline{\chi }$
\begin{equation}
\overline{\chi }=\chi _{eff}-(1-\phi)\frac{\partial \chi
_{eff}}{\partial \phi } \label{chib}
\end{equation}
rather then on $\chi $. Since $\mu _{s}$ determines the
colligative properties of the solution, measurements of such
properties yield $\overline{\chi }$ rather than $\chi _{eff}$. It
is the $\overline{\chi }$ values that are usually reported in the
literature. Power series in $\phi $ provide a useful
representation of the experimentally tabulated $\overline{\chi }$
values\cite{wolf}
\begin{equation}
\overline{\chi }(T,\phi)=\sum_{i=0}^{n}\overline{\chi }_{i}(T)\phi
^{i} \label{chic}
\end{equation}
Typically, it is sufficient to utilize the first three terms in
this expansion, that is, $\overline{\chi }$ can be well fitted by
$\overline{\chi }(T,\phi)=$ $\overline{\chi }_{0}+\overline{\chi
}_{1}\phi +\overline{\chi } _{2}\phi ^{2}$ where $\chi _{0}$ is
often close to $1/2$ and all coefficients are, in principle, $T$
dependent. From the measured $\overline{ \chi }(\phi)$ it is
possible to obtain $\chi _{eff}(\phi)$ up to an additive constant

\begin{equation}
\chi _{eff}(\phi)=\frac{\chi _{eff}(0)-\int_{0}^{\phi
}\overline{\chi } (\phi)d\phi }{1-\phi },
\end{equation}
where the integration constant, $\chi_{eff}(0)$, is the value $\chi_{eff}$
at $\phi=0$.\cite{chiefff}

The replacement of $\chi (T)$ by $\chi _{eff}(T,\phi)$ can result
in qualitative change in the phase behavior of the polymer
solutions.\cite{n-Cluster,Qian,Solc1,Solc2} In the following we
focus, following de Gennes,\cite{n-Cluster} on the limit of
$N\rightarrow \infty $ when the novel features of the phase
behavior are simple to discern. Importantly, this is the limit
relevant to brushes of grafted chains of finite $N$ because the
anchoring freezes out the translational degrees of freedom of the
chains. In the familiar case, of $\chi (T)$ and $N\rightarrow
\infty $, the resulting phase separation involves a coexistence of
a concentrated polymer solution with a {\em pure} solvent. In
marked contrast, solutions characterized by $ \chi _{eff}(T,\phi
)$ can exhibit, in the $N\rightarrow \infty $ limit, a second type
of phase separation. This involves a coexistence of two phases of
{\em non-zero} polymer concentration. This last feature is a
necessary ingredient for the occurrence of a vertical phase
separation within a brush. The features noted above can be
discerned from the critical points of the solution, as specified
by $\partial ^{2}f(\phi)/\partial \phi ^{2}=\partial ^{3}f(\phi
)/\partial \phi ^{3}=0$ or

\begin{equation}
\left\{
\begin{array}{l}
\frac{1}{N\phi_{c}}+\frac{1}{1-\phi _{c}}-2\overline{\chi }(\phi
_{c})-\phi_{c}\left. \frac{\partial \overline{\chi }(\phi
)}{\partial \phi }\right|_{\phi _{c}}=0 \\
-\frac{1}{N\phi _{c}^{2}}+\frac{1}{\left( 1-\phi _{c}\right)
^{2}}-3\left. \frac{\partial \overline{\chi }(\phi)}{\partial \phi
}\right| _{\phi _{c}}-\phi _{c}\left. \frac{\partial
^{2}\overline{\chi }(\phi)}{\partial \phi ^{2}}\right| _{\phi
_{c}}=0
\end{array}
\right. \label{CritP}
\end{equation}
In the case of $\chi (T)$ these lead to the familiar critical
point specified by $\phi _{c}=1/(1+\sqrt{N})$ and $\chi
_{c}=\frac{1}{2}(1+\frac{1 }{\sqrt{N}})^{2}$. Accordingly, in the
limit of $N\rightarrow \infty $ we have $\phi _{c}\rightarrow 0$
and $\chi _{c}\rightarrow 1/2$. When $\chi (T)$ is replaced by
$\chi _{eff}(T,\phi)$ an additional critical point, associated
with the second type of phase separation, emerges. For
illustration purpose we will consider the $N\rightarrow \infty $
limit for the case of $\overline{\chi }(\phi)=1/2+\overline{\chi
}_{2}(T)\phi ^{2}$ when equation (\ref{CritP}) yields an extra
critical point at $\phi _{c}=1/2$ and $\overline{\chi }_{2c}=1$
{\em i.e.}, for $\overline{\chi }_{2}>1$ the system undergoes
phase separation involving the coexistence of a dilute phase of
concentration $\phi_{-}>0$, and a dense phase of concentration $
\phi _{+}>\phi _{-}$.\cite{foot} At the vicinity of the critical
point the binodal is well approximated by the spinodal curve
$\partial ^{2}f(\phi)/\partial \phi ^{2}=0$. In the limit
$N\rightarrow \infty $ the spinodal is specified by

\begin{equation}
\frac{1}{1-\phi }-1-4\overline{\chi }_{2}\phi ^{2}=0.
\label{spinodal}
\end{equation}
and the approximate values of $\phi _{+}$ and $\phi _{-}$ are

\begin{equation}
\phi _{\pm }=\frac{1}{2}\pm
\frac{1}{2}\sqrt{1-\frac{1}{\overline{\chi } _{2}(T)}}.
\end{equation}

Our discussion thus far concerned the thermodynamics of solutions
of free polymers, when the translational degrees of freedom of the
chains play a role. In the following, we mostly focus on the
swelling behavior of free isolated chains and of brushes immersed
in a pure solvent. In these situations $f$ as discussed above is
replaced by $f_{\infty }$ corresponding to the limit of
$N\rightarrow \infty $
\begin{equation}
\frac{f_{\infty }}{kT}=(1-\phi)\ln (1-\phi)+\chi _{eff}(\phi)\phi
(1-\phi). \label{Fmix}
\end{equation}
$f_{\infty }$ determines the swelling behavior of brushes because the
terminally anchored chains lose their translational entropy. Similarly, the
swelling of isolated free coils is controlled by $f_{\infty }$ because the
motion of the chain center of mass is irrelevant to this process. In this
last case it is important to note that $\phi $ refers to the monomer
concentration within the coil rather than to the average concentration of
the solution. The osmotic pressure corresponding to $f_{\infty }$ is
\begin{equation}
\frac{\pi _{\infty }a^{3}}{kT}=-\phi -\ln (1-\phi)-\overline{\chi
}(\phi)\phi ^{2} \label{Pio}
\end{equation}
To gain insight concerning the significance of $\chi _{eff}(\phi)$
it is helpful to consider the expansions of $f_{\infty }$ and $\pi
_{\infty }$. Two routes are of interest. In the first we follow
the procedure adopted in the standard discussions involving $\chi
(T)$ and replace the logarithmic term in (\ref{Fmix}) by its
series expansion leading to
\begin{equation}
\frac{f_{\infty }}{kT}=(\chi _{eff}-1)\phi
+\frac{1}{2}\widehat{v}\phi ^{2}+ \frac{1}{6}\phi ^{3}+...
\label{fv}
\end{equation}
Here the ``excluded volume parameter''
$\widehat{v}(T,\phi)=1-2\chi _{eff}(T,\phi)$ is dependent on both
$T$ and $\phi $. Note, that in this case it is important to retain
the ``linear'' term because of the $\phi $ dependence of $\chi
_{eff}$. The corresponding expression for $\pi _{\infty } $ is
\begin{equation}
\frac{\pi _{\infty }a^{3}}{kT}\approx \frac{1}{2}\overline{v}\phi
^{2}+\frac{ 1}{3}\phi ^{3}+\frac{1}{4}\phi ^{4}+..., \label{piv}
\end{equation}
where $\overline{v}(T,\phi)=1-2\overline{\chi }(T,\phi)\neq
\widehat{v}$ is, again dependent on both $T$ and $\phi $. Thus
when following this route the effect of replacing $\chi (T)$ by
$\chi _{eff}(T,\phi)$ is two fold: (i) the coefficients of $\phi
^{2}$ in the expansions of $f_{\infty }$ and of $\pi _{\infty }$
become $\phi $ dependent (ii) The coefficients of $\phi ^{2}$ in
the expansions of $f_{\infty }$ and of $\pi _{\infty }$ are
different. All coefficients of higher order terms are positive
constants. Clearly, this is also the case for $f$ and for $\pi $.
The second route is to replace $\chi _{eff}(T,\phi)$ and
$\overline{\chi }(T,\phi)$ by their power series in $\phi $. As
was noted earlier, the power series of $\chi _{eff}(T,\phi)$ is
specified by the one for $\overline{\chi }(T,\phi)$. Yet, for the
purpose of our discussion it is sufficient to use $\chi
_{eff}(T,\phi)=\sum_{i=0}^{n}\widehat{\chi }_{i}(T)\phi ^{i}$
without specifying the relationship between $\widehat{\chi
}_{i}(T)$ and $\overline{ \chi }_{i}(T)$. Following this second
route we obtain

\begin{eqnarray}
\frac{f_{\infty }}{kT}&=&(\widehat{\chi }_{0}-1)\phi
+(\frac{1}{2}-\widehat{ \chi }_{0}+\widehat{\chi }_{1})\phi
^{2}+\nonumber \\
&&(\frac{1}{6}-\widehat{\chi }_{1}+ \widehat{\chi
}_{2})\phi ^{3}+...
\end{eqnarray}

\begin{equation}
\frac{\pi _{\infty }a^{3}}{kT}\approx
(\frac{1}{2}-\overline{\chi}_{0})\phi
^{2}+(\frac{1}{3}-\overline{\chi }_{1})\phi
^{3}+(\frac{1}{4}-\overline{\chi }_{2})\phi ^{4}+...,
\end{equation}
In this form none of the coefficients depend on $\phi $ but all
are $T$ dependent and capable of changing sign. In marked
distinction, when $\chi (T) $ only the first coefficient is $T$
dependent and capable of change of sign. The role of $v=1-2\chi
(T)$ as a measure of solvent quality is traceable to this last
quality. The expansions (\ref{fv}) and (\ref{piv}) retain this
characteristic at the price of introducing a $\phi$ dependence of
$\overline{v}$ and $\widehat{v}$. As we shall discuss
$\overline{\chi } (T,\phi)$ is the counterpart of $\chi(T)$ as an
indicator of the solvent quality. Accordingly, $\overline{v}$
affords some the usefulness of $ v=1-2\chi (T)$.

Thus far our discussion concerned the thermodynamics of polymer
solutions characterized by $\chi _{eff}(T,\phi)$. The microscopic
swelling behavior of coils and brushes, when modelled as systems
of uniform concentration, is described respectively by the Flory
and Alexander approximations. Within these approximations the
swelling behavior reflects a balance between the osmotic pressure
$\pi _{\infty }$ and the elastic restoring force. In the Flory
approximation an isolated coil is viewed as a sphere of radius $R$
with a uniform monomer density $\phi \approx Na^{3}/R^{3}$ where
$a$ is the monomer size. Within the Alexander model one considers
a planar brush of terminally anchored chains such that the
grafting density is uniform and the area per chain, $\sigma $, is
constant. The grafting density is high so as to enforce chain
crowding, $\sigma \leq R_{F}^{2}$ where $R_{F}\approx N^{3/5}a$ is
the Flory radius of the isolated coil. The brush is considered as
a planar layer of thickness $R$ and uniform density $\phi \approx
Na^{3}/\sigma R$.

The free energy per chain, in both cases, is

\begin{equation}
F_{chain}=f_{\infty }V_{chain}/a^{3}+F_{el} \label{Fchain}
\end{equation}
where $f_{\infty }$ is the mixing free energy per lattice site, $V_{chain}$
is the volume per chain and $F_{el}$ is its elastic free energy. In both
cases the elastic free energy is approximated by

\begin{equation}
\frac{F_{el}}{kT}\approx \frac{R^{2}}{R_{0}^{2}}+\frac{R_{0}^{2}}{R^{2}},
\label{Fel}
\end{equation}
where $R_{0}\approx N^{1/2}a$ is the radius of an ideal, Gaussian coil. The
free energies of the coil and the brush differ because of $V_{chain}$

\begin{equation}
V_{chain}\approx \left\{
\begin{array}{l}
R^{3}\qquad \text{coil} \\
R\sigma \qquad \text{brush}
\end{array}
\right. . \label{Vchain}
\end{equation}
In turn, this reflects the different geometries of the two
systems. The coil is spherical while the brush is planar. The
swelling behavior is specified by the equilibrium condition
$\partial F_{chain}/\partial R=0$. Since $
\partial R\approx -(R/\phi)\partial \phi $ this leads to
\begin{equation}
\frac{\partial F_{el}}{\partial R}\approx \pi _{\infty
}\frac{dV_{chain}}{dR} \approx \left\{
\begin{array}{l}
\pi _{\infty }R^{2}\qquad \text{coil} \\
\pi _{\infty }\sigma \qquad \text{ brush}
\end{array}
\right. .
\end{equation}
In the next two sections we will analyze the consequences of this
equation for polymeric systems with $\chi_{eff}(\phi)$.

Before we proceed, a note of caution. The program outlined above
calls for utilizing $\overline{\chi }(\phi)$, as obtained from
thermodynamic measurements, to determine the swelling behavior of
coils and brushes. It is based on the assumption that the measured
$\overline{\chi }(\phi)$ is identical to the one experienced by
the coils. This is non trivial assumption since in certain
models\cite{Painter} the $\phi $ dependence of $ \overline{\chi
}(\phi)$ arises because of an interplay of intra and interchain
contacts. Within such models the $\overline{\chi }(\phi)$
experienced by a coil may differ from the measured $\overline{\chi
}(\phi)$.

\section{Coils -- Swelling and Collapse within the Flory Approximation with $
\protect\chi_{eff}(\protect\phi)$}

As noted in the introduction, the swelling behavior of an isolated
coil is an important signature of the solvent quality. When $\chi
_{eff}(T,\phi)$ replaces $\chi (T)$ the swelling behavior is
modified. Two features are of special interest. One concerns the
locus of the cross-over, $g_{B}$, between the $N^{1/2}$ and
$N^{3/5}$ scaling in the ``nearly good solvent'' regime. Chains
with $N<g_{B}$ exhibit ideal coil behavior while longer chains
exhibit self avoidance statistics\cite{PGG} and their span scales
as $ N^{3/5} $. The $\phi $ dependence of $\overline{\chi }(\phi
)$ can result in a significant shift in $g_{B}$. A second,
qualitative, effect concerns the collapse transition within the
Flory approximation. When $\overline{\chi } (\phi)$ increases with
$\phi $ the collapse can assume the character of a first-order
phase transition.

In good solvent conditions, when $\phi \ll 1$ and $R>N^{1/2}a$,
only the first term in the elastic free energy (\ref{Fel}) plays a
role. The equilibrium condition for a coil, $\partial
F_{el}/\partial R\approx \pi _{\infty }R^{2}$, reduces to
$R/Na^{2}\approx (\pi _{\infty }/kT)R^{2}$. Since $\phi \ll 1$ we
retain only the first two terms in $\overline{\chi } (\phi
)\approx \overline{\chi }_{0}+\overline{\chi }_{1}\phi $, and
obtain $ \pi _{\infty }a^{3}/kT\approx
\frac{1}{2}(1-2\overline{\chi }_{0})\phi ^{2}+(
\frac{1}{3}-\overline{\chi }_{1})\phi ^{3}$. $R\sim N^{3/5}$
scaling is obtained when the first term in the expansion for $\pi
_{\infty }$ is dominant while $R\sim N^{1/2}$ is found when the
second term dominates. The cross-over between the two regimes
occurs when $\phi _{B}\approx (1-2 \overline{\chi
}_{0})/(\frac{1}{3}-\overline{\chi }_{1})$. Identifying $\phi
_{B}\approx g_{B}a^{3}/r_{B}^{3}$, where $r_{B}\approx
(\frac{1}{3}- \overline{\chi }_{1})^{1/8}g_{B}^{1/2}a$ is the span
of the corresponding ideal chain, leads to

\begin{equation}
g_{B}\approx \frac{\left( \frac{1}{3}-\overline{\chi }_{1}\right)
^{5/4}}{ \left( 1-2\overline{\chi }_{0}\right) ^{2}}
\label{Ntilde}
\end{equation}
where both $\chi _{1}$ and $\chi _{0}$ are $T$ dependent. This expression
for $g_{B}$ reduces to the familiar one, as obtained for solutions
characterized by $\chi (T)$, when $\chi _{1}=0$. The two asymptotic regimes
for the radius of the chain are

\begin{equation}
\frac{R}{a}\approx \left\{
\begin{array}{l}
\left( \frac{1}{3}-\overline{\chi }_{1}\right) ^{1/8}N^{1/2}\text{
\qquad }
N\ll g_{B} \\
\left( \frac{1}{2}-\overline{\chi }_{0}\right) ^{1/5}N^{3/5}\qquad
\text{ } N\gg g_{B}
\end{array}
\right.
\end{equation}
As noted earlier, the effect on $g_{B}$ can be significant. Thus
for polystyrene in toluene at 25$^{0}C$ $\overline{\chi
}_{0}=0.431$ and $ \overline{\chi }_{1}=-0.311$\cite{wolf} and
$g_{B}\approx 120$ as compared to $g_{B}\approx 50$ obtained when
the $\phi $ dependence is neglected {\em i.e.}, $\chi _{1}=0$.

\begin{figure}[h]
\begin{center}
\epsfig{file=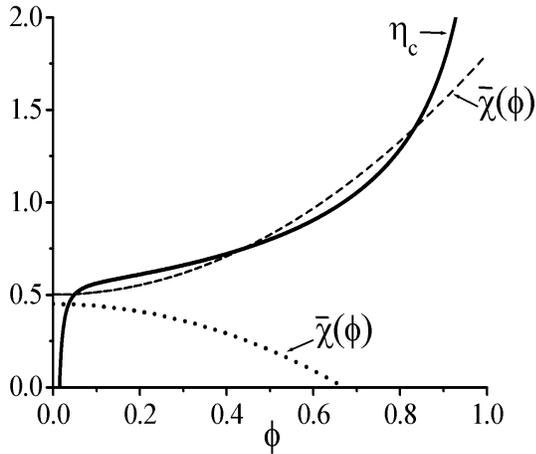, width=7cm, height=6cm} \caption{Graphical
solution of the equilibrium condition of an isolated coil
($N=300$) depicting the crossing of $\eta _{c}(N,\phi)$, as given
by (\ref{equat}) with $\overline{\chi }(\phi)$. The two curves
cross once when $\overline{\chi}(\phi)$ decreases with $\phi $
(dots) but three intersections, indicating a first order phase
transition, may occur when $\overline{\chi}(\phi)$ increases with
$\phi$ (dashes).}
\end{center}
\end{figure}

When the $\phi$ dependence of $\chi$ is overlooked it is
convenient to obtain $g_{B}$ in terms of a perturbative parameter
$\zeta$ measuring the relative importance of the repulsive binary
interactions as compared to an elastic energy of order $kT$.
Within this approach $g_{B}$ corresponds to $ \zeta =1$.\cite{PGG}
This approach can not however account for the contribution of $T$
dependent higher order terms.

An analytical solution of $\partial F_{el}/\partial R=\pi _{\infty
}R^{2}$ for arbitrary $\overline{\chi }(\phi)$ is not feasible.
Yet, one may gain insight concerning the swelling and collapse
behavior from the general features of the graphical solution of
the equilibrium condition. Neglecting numerical factors $\partial
F_{el}/\partial R=\pi _{\infty }R^{2}$ can be written as
\begin{eqnarray}
\overline{\chi}(\phi)&\approx&-\frac{\ln (1-\phi)+\phi }{\phi
^{2}}+\nonumber \\
&&\frac{1}{N^{2/3}\phi^{1/3}}-\frac{1}{N^{4/3}\phi ^{5/3}}\equiv
\eta _{c}(\phi,N) \label{equat}
\end{eqnarray}

The equilibrium states correspond to the intersections of
$\overline{\chi } (\phi)$ and $\eta _{c}(\phi,N)$ (Figure 1). When
$\chi _{eff}(\phi)= \overline{\chi }(\phi)=\chi $ the two curves
intersect at one point only. This is also the case when
$\overline{\chi }(\phi)$ is a decreasing function of $\phi $.
However, in this situation the intersection occurs at lower $\phi
$ in comparison to the intersection between $\eta _{c}(\phi,N)$
and $\overline{\chi }(\phi)=\chi _{0}$ thus indicating stronger
swelling. When $\overline{\chi }(\phi)$ is an increasing function
of $\phi $ it is possible to distinguish between two important
scenarios: (i) $\overline{\chi }(\phi)$ and $\eta _{c}(\phi)$
intersect at a single point. In comparison to the intersection of
$\eta _{c}(\phi,N)$ and $\overline{\chi }(\phi)=\chi _{0}$, this
occurs at higher $\phi $ thus indicating weaker swelling. (ii)
$\overline{\chi }(\phi)$ and $\eta _{c}(\phi)$ intersect at three
points. This case corresponds to a $F_{chain}$ (\ref{Fchain})
exhibiting two minima separated by a maximum thus indicating a
collapse taking place as a first order phase transition. Note that
this last scenario occurs only for $ \overline{\chi }(\phi)$ that
increases with $\phi $. Arguing that higher $ \overline{\chi
}(\phi)$ values indicate a poorer solvent allows for a simple
interpretation of this result. When $R$ shrinks, $\phi $ increases
leading to a higher $\overline{\chi }(\phi)$. Accordingly, the
effective solvent quality diminishes with $R$ thus giving rise to
cooperativity leading to a first-order collapse transition. At
this point it is important to stress the limitations of the Flory
approach as described above. Since the monomer volume fraction,
$\phi$, is assumed to be uniform, this model does not allow for
the possibility of radial phase separation within a single
globule. To investigate such scenarios it is necessary to utilize
the Lifshitz theory of collapse.\cite{Lif} This however is beyond
the scope of this work.

While the character of the collapse transition within the Flory
approximation is affected by the $\phi $ dependence of $\chi
_{eff}$, there is essentially no change in span of the collapsed
chain as specified by the condition $\pi_{\infty }=0$. When $\chi
(T)$ is independent of $\phi $ and $ \phi \ll 1$ this condition
leads to $v\phi ^{2}\sim \phi ^{3}$ and to $R/a\sim
|v|^{-1/3}N^{1/3}$. For concentration dependent $\chi
_{eff}(T,\phi) $, $\pi _{\infty }a^{3}/kT\approx
\frac{1}{2}(1-2\overline{\chi }_{0})\phi
^{2}+(\frac{1}{3}-\overline{\chi }_{1}\phi ^{3})=0$, leads to

\begin{equation}
\frac{R}{a}\approx \left( \frac{\frac{1}{3}-\overline{\chi
}_{1}}{\overline{ \chi }_{0}-\frac{1}{2}}\right) ^{1/3}N^{1/3}
\end{equation}
Thus $R\sim N^{1/3}$ is retained but with a modified numerical prefactor and
an additional $T$ dependence introduced by $\overline{\chi }_{1}$.

\section{Brushes -- Swelling and Collapse within the Alexander Approximation
with $\protect\chi _{eff}(\protect\phi)$}

The swelling behavior of a brush within the Alexander model
exhibits similar trends to those found in the case of the isolated
coil. In a good solvent, when $\phi \ll 1$ and $R>N^{1/2}a$, the
equilibrium condition for the brush, $\partial F_{el}/\partial
R\approx \pi _{\infty }\sigma $, leads to $R/Na^{2}\approx \left(
\pi _{\infty }/kT\right) \sigma $ or $R/a\approx N(\pi _{\infty
}a^{3}/kT)(\sigma /a^{2})$. As in the case of the coil, the
``nearly good solvent'' case involves two regimes. The cross-over
occurs at
\begin{equation}
\sigma _{B}\approx r_{B}^{2}\approx \left(
\frac{1}{3}-\overline{\chi }_{1}\right) ^{1/4}g_{B}a^{2}
\label{brush1}
\end{equation}
When $\sigma >\sigma _{B}$ the chains in the brush exhibit self-avoidance
and $R/a\approx N\left( a^{2}/\sigma \right) ^{1/3}$ while for $\sigma
<\sigma _{B}$ one obtains $R/a\approx N\left( a^{2}/\sigma \right) ^{1/2}$
corresponding to a brush of ideal chains.

\begin{figure}[h]
\begin{center}
\epsfig{file=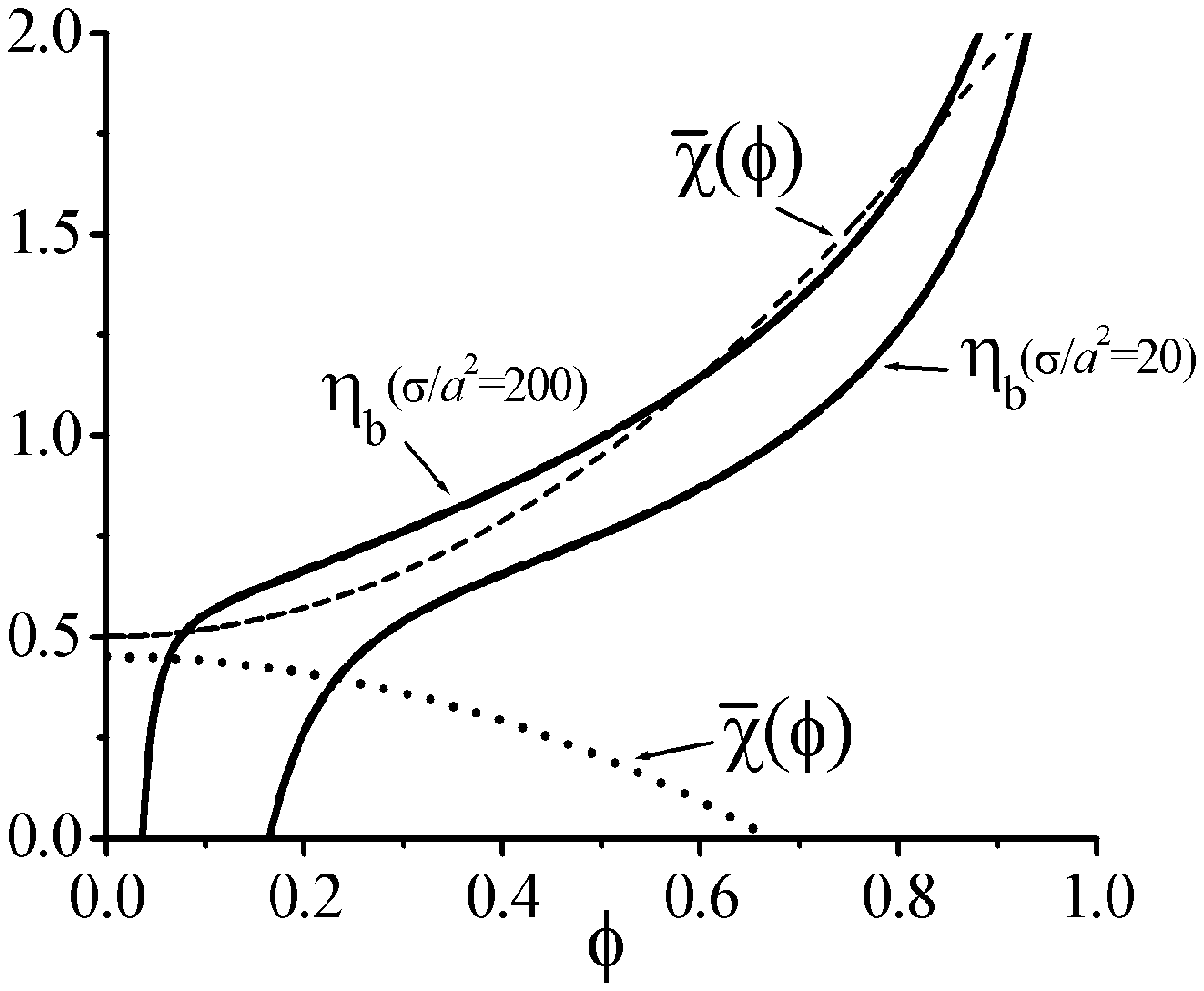, width=7cm, height=6cm} \caption{Graphical
solution of the equilibrium condition of a brush ($N=300$)
depicting the crossing of $\eta_{b}(N,\phi ,\sigma)$, as given by
(\ref{brush3}) with $\overline{\chi }(\phi)$. The two curves cross
once when $\overline{\chi }(\phi)$ decreases with $\phi $ (dots)
but three intersections, indicating a first order phase
transition, may occur when $\overline{\chi }(\phi)$ increases with
$\phi $ (dashes).}
\end{center}
\end{figure}

As in the case of a collapsed globule, the thickness of the fully collapsed
brush is essentially unaffected by the $\phi $ dependence of $\chi _{eff}$.
The thickness is determined by $\pi _{\infty }=0$ but with $\phi \approx
Na^{3}/\sigma R$ rather than $\phi \approx Na^{3}/R^{3}$ leading to
\begin{equation}
\frac{R}{a}\approx \left( \frac{\frac{1}{3}-\overline{\chi
}_{1}}{\overline{ \chi }_{0}-\frac{1}{2}}\right)
N\frac{a^{2}}{\sigma }. \label{brush2}
\end{equation}
that is, the $R\sim N(a^{2}/\sigma)$ scaling is retained with a
modified prefactor and an additional $T$ dependence due to $\chi
_{1}$.

Again, an analytical solution of $\partial F_{el}/\partial R=\pi
_{\infty }\sigma $ for arbitrary $\overline{\chi }(\phi)$ is not
practical but it is of interest to consider the graphical solution
of the equilibrium condition $\partial F_{el}/\partial R=\pi
_{\infty }\sigma $ expressed as

\begin{eqnarray}
\overline{\chi }(\phi) &\approx & -\frac{\ln (1-\phi)+\phi }{\phi
^{2}}- \nonumber \\
&&\frac{1}{\phi ^{3}}\left( \frac{a^{2}}{\sigma }\right)
^{2}+\frac{\phi }{N^{2}} \left( \frac{\sigma }{a^{2}}\right)
^{2}\equiv \eta _{b}(\phi,N,\sigma) \label{brush3}
\end{eqnarray}

The equilibrium state is specified by the intersections of
$\overline{\chi } (\phi)$ and $\eta _{b}(\phi,N,\sigma)$ (Figure
2). When $\chi _{eff}(\phi)=\overline{\chi }(\phi)=\chi $ the two
curves intersect at one point. Similar behavior is found when
$\overline{\chi }(\phi)$ is a decreasing function of $\phi $. In
comparison to the intersection between $\eta _{b}(\phi,\sigma)$
and $\overline{\chi }(\phi)=\overline{\chi }_{0}$ the intersection
occurs at lower $\phi $ thus signaling stronger swelling. As in
the case of the coil, it is possible to distinguish between two
important scenarios when $\overline{\chi }(\phi)$ increases with
$\phi $: (i) $ \overline{\chi }(\phi)$ and $\eta _{b}(\phi
,N,\sigma)$ intersect at a single point. This occurs at higher
$\phi $ in comparison to the intersection of $\eta _{b}(\phi,N)$
and $\overline{\chi }(\phi)=\overline{ \chi }_{0}$ and corresponds
to weaker swelling. (ii) $\overline{\chi }(\phi) $ and $\eta
_{b}(\phi,N,\sigma)$ intersect at three points. In this case
$F_{chain}$ exhibits two minima separated by a maximum and the
collapse takes place as a first-order phase transition. As we
shall discuss shortly, a first order ``collapse transition'' is
indeed possible when $\overline{ \chi }(\phi)$ increases with
$\phi $. However, this transition involves a vertical phase
separation within the brush. To properly analyze this case it is
necessary to allow for the spatial variation of $\phi $ thus
requiring a more refined description of the brush. This topic is
addressed in the next section.

\section{Brushes -- Swelling and Collapse within the Pincus Approximation
with $\protect\chi _{eff}(\protect\phi)$}

The discussion of the two preceding sections concerned the
``global'' solvent quality in systems of assumed uniform density.
To explore the coupling of the ``local'' solvent quality with a
spatially varying $\phi $ we reanalyze the swelling behavior of a
brush using the Pincus approximation instead of the Alexander
model. The Alexander model invokes two assumptions: (i) uniform
density that is, $\phi $ behaves as a step function thus endowing
the brush with a sharp boundary. (ii) The chains are uniformly
stretched with their ends straddling the sharp boundary of the
brush. While this approximation allows to recover the correct
scaling behavior of the brush, the two underlying assumption are
in fact wrong. Both $\phi $ and the local extension of the chains
vary with the distance from the grafting surface, $z$, and the
chain ends are distributed throughout the layer. The Self
Consistent Field (SCF) theory of brushes furnishes a rigorous
description of these features.\cite{Milner,Zhulina} The SCF theory
provides a basis for the analysis of the coupling of $\chi
_{eff}(\phi)$ and $\phi (z)$. Indeed, such analysis was already
carried out for the special case of brushes described by the
$n$-cluster model. In the following we will use a simpler scheme
proposed by Pincus.\cite{Pincus,AHcluster} The level of this
approximation is roughly midway between the Alexander model and
the SCF theory. It retains the uniform stretching assumption but
allows for spatial variation in $\phi $ and in the distribution of
the ends. Within this approach the free energy per unit area of
the brush is $\gamma =a^{-3}\int_{0}^{R}f_{brush}dz$, where
$f_{brush}$ is the corresponding free energy density
\begin{equation}
\frac{f_{brush}}{kT}=\frac{f_{\infty
}}{kT}+\frac{z^{2}}{Na^{2}}\Psi (z)-\lambda \phi (z)
\label{Fbrush}
\end{equation}
The second term allows for the elastic free energy of the chains.
A chain having an end at altitude $z$ is assumed to be uniformly
stretched and is thus allocated an elastic penalty of
$F_{el}/kT\approx z^{2}/Na^{2}$. The chains' ends are assumed to
be distributed throughout the layer with a volume fraction $\Psi
(z)$.

\begin{figure}[h]
\begin{center}
\epsfig{file=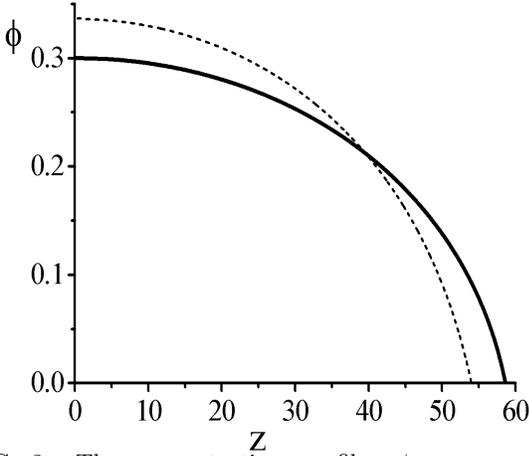, width=7cm, height=6cm} \caption{The
concentration profile, $\phi $ {\em vs.} $z$ plots, for a brush of
polystyrene in toluene with $N=300$, $\sigma /a^{2}=22.5$ ($
\sigma /R_{F}^{2}=0.024$) $T=25^{0}C$ calculated using
$\overline{\chi } (\phi)=0.431-0.311\phi -0.036\phi
^{2}$.\cite{wolf} The dashed line depicts the case $\overline{\chi
}(\phi)=\overline{\chi}_{0}$ while the continuous line describes
$\phi (z)$ for the full $\overline{\chi }(\phi)$.}
\end{center}
\end{figure}

The core of the Pincus approximation is the assumption that local
concentration of ends scales as the fraction of ends within the
chain, $1/N$, that is
\begin{equation}
\Psi (z)=\frac{\phi (z)}{N} \label{Psi}
\end{equation}

As opposed to the SCF theory, $\Psi (z)$ is assumed and not
derived. While $ \Psi (z)$ is wrong for small altitudes the
approximation yield the correct $ \phi (z)$ because $F_{el}\approx
z^{2}$ and the large $z$ contribution, where the assumed $\Psi
(z)$ is reasonable, dominates. Finally, $\lambda $ is a Lagrange
parameter fixing the number of monomers per chain, $N$.

The equilibrium concentration profile $\phi (z)$ is specified by the
condition $\delta \gamma /\delta \phi =0$. Since $f_{brush}$ does not depend
on $d\phi /dz$ the equilibrium condition is $\delta \gamma /\delta \phi
=\partial f_{brush}/\partial \phi =0$ or

\begin{equation}
\mu _{\infty }(\phi)/kT=\lambda -Bz^{2}, \label{profile}
\end{equation}
Here $\mu _{\infty }=\partial f_{\infty }/\partial \phi $ is the exchange
chemical potential as obtained from (\ref{Fmix})\cite{muexp}
\begin{equation}
\mu _{\infty }(\phi)/kT=-\ln (1-\phi)-1+\chi _{eff}(\phi)\left(
1-\phi \right) -\phi \overline{\chi }(\phi).
\end{equation}
In the following we utilize $B=3\pi^{2}/8N^{2}a^{2}$, as obtained
from the SCF theory, rather than the value obtained from the
Pincus model. Upon making this substitution, equation
(\ref{profile}) is identical to the one obtained from the rigorous
SCF theory. In the following we impose the condition $\phi
_{R}\equiv \phi (R)=0$. This condition is sufficient for our
discussion since we are interested in the case of a vertical phase
separation within a brush due to a ``second type'' of phase
separation involving coexistence of two phases of finite
concentration.

\begin{figure}[h]
\begin{center}
\epsfig{file=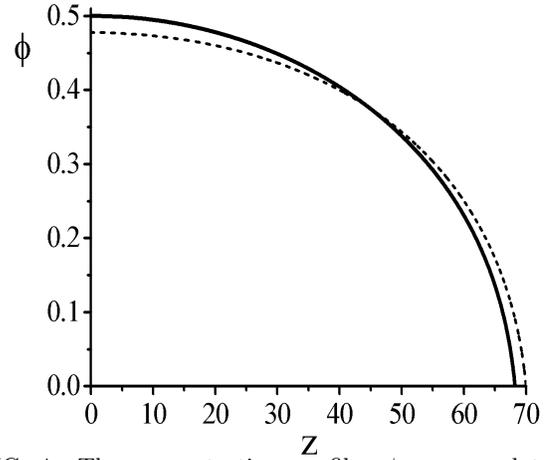, width=7cm, height=6cm} \caption{The
concentration profile, $\phi$ {\em vs}. $z$ plots, for a brush of
PEO in water with $N=300$, $\sigma/a^{2}=11$ ($\sigma
/R_{F}^{2}=0.012$), $T=68^{0}C$ calculated using $\overline{\chi
}(\phi)=0.469+$
$0.060\phi-0.241\phi^{2}+0.370\phi^{3}+0.579\phi^{4}$ obtained by
polynomial fit of the $\overline{\chi}(\phi)$ data in Ref
\cite{Molyneux}. The dashed line depicts the case
$\overline{\chi}(\phi)=\chi_{0}$ while the continuous line
describes $\phi(z)$ for the full $\overline{\chi}(\phi)$.}
\end{center}
\end{figure}

In the general case, the condition $\phi _{R}=0$ is replaced by
$\pi _{\infty }(\phi _{R})=0 $, thus allowing for a fully
collapsed brush where $\phi _{R}>0$. Since in our case $\phi
_{R}=0$, equation (\ref{profile}) specifies $\lambda $
\begin{equation}
\lambda =BR^{2}+\mu _{\infty }(0)/kT. \label{lambda}
\end{equation}
thus enabling us to rewrite (\ref{profile}) as
\begin{equation}
\Delta \mu _{\infty }(\phi)/kT=B(R^{2}-z^{2}) \label{Nprofle}
\end{equation}
where $\Delta \mu _{\infty }(\phi)\equiv \mu _{\infty }(\phi)-\mu
_{\infty }(0)$ or $\Delta \mu _{\infty }(\phi)/kT=\mu _{\infty
}(\phi)/kT+1-\chi _{eff}(0)$. Equation (\ref{Nprofle}) determines
the height $R$ of the brush for a given $\phi (z=0)\equiv \phi
_{0}$
\begin{equation}
R=\sqrt{\Delta \mu _{\infty }(\phi _{0})/BkT} \label{height}
\end{equation}
and the concentration profile in the form
\begin{equation}
z(\phi)=\sqrt{\left[ \Delta \mu _{\infty }(\phi_{0})-\Delta \mu
_{\infty }(\phi)\right] /BkT.} \label{Fprofile}
\end{equation}
The grafting density corresponding to $\phi _{0}$ is then specified by the
constraint
\begin{equation}
\frac{Na^{3}}{\sigma }=\int_{0}^{R}\phi (z)dz \label{ConfN1}
\end{equation}

When $\chi (T)$ and $f_{\infty }$ is approximated by $f_{\infty
}\approx v(T)\phi ^{2}$ the concentration profile of the brush is
parabolic. Upon replacing $\chi (T)$ by $\chi _{eff}(T,\phi)$ the
concentration profile of the brush $\phi (z)$ is modified because
the solvent quality varies, in effect, with the altitude $z$.

\begin{figure}[h]
\begin{center}
\epsfig{file=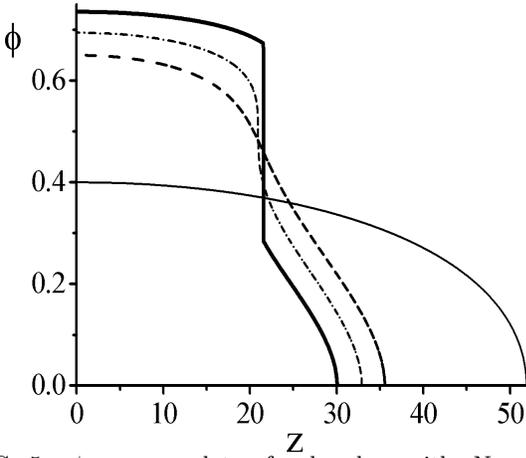, width=7cm, height=6cm} \caption{$\phi$ {\em
vs}. $z$ plots, for brushes with $N=300$, $\sigma /a^{2}=18$
($\sigma /R_{F}^{2}=0.019$) subject to interactions described by
$\overline{\chi }(\phi)=0.5$ (thin line), $\overline{\chi }
(\phi)=0.5+$ $0.95\phi ^{2}$ (dashes), $\overline{\chi
}(\phi)=0.5+$ $ 1.00\phi ^{2}$(dash-dots), $\overline{\chi
}(\phi)=0.5+$ $1.05\phi ^{2}$ (thick line).}
\end{center}
\end{figure}

As a result, $\phi (z)$ is no longer parabolic. Three principle
scenarios are possible. When $\overline{\chi} (\phi)$ is a
decreasing function of $\phi $ the brush height, $R$, increases
while the concentration at the grafting surface, $\phi _{0}$,
decreases. Such behavior is expected, for example, for brushes of
polystyrene in toluene (Figure 3). When $\overline{\chi }(\phi)$
is an increasing function of $\phi $ two scenarios are of
interest. When there is no phase separation of the second type,
the qualitative features of $\phi (z) $ are not modified. However,
the brush height, $R$ decreases while the concentration at the
grafting surface, $\phi _{0}$ increases. Such is the case for
polyetheleneoxide (PEO) brushes in water (Figure 4). A
qualitatively different scenario occurs when a second type of
phase separation occurs within the brush leading to a
discontinuity in $\phi (z)$. This case was first discussed by
Wagner {\em et al} in the context of the {\em n}-cluster
model.\cite{Wagner,AHcluster} When $\overline{\chi }(\phi)$
increases with $\phi $ to the extent that a bulk phase separation
of the second type occurs it is possible to distinguish between
two regimes. For grafting densities lower than $\sigma _{c}$, to
be specified below, $\phi _{0}<\phi _{-}$. In this range $\phi
(z)<\phi _{-}$ at all altitudes and there is no phase separation.
On the other hand, when $\sigma >\sigma _{c}$ phase separation
occurs within the brush leading to a discontinuous $\phi(z)$.

The onset of phase separation within the brush for $\sigma >\sigma
_{c}$ is signalled by the appearance of multiple roots to equation
(\ref{Fprofile}). These are due to the van der Waals loop traced
by $\mu _{\infty }(\phi)$ in the range $\phi _{-}<\phi <\phi
_{+}$. The corresponding concave region in $f_{\infty }$ gives
rise to an unstable domain in $f_{brush}$. The coexistence within
the brush is specified by two conditions: (i) $\mu _{brush}(\phi
_{+})=\mu _{brush}(\phi _{-})$ where $\mu _{brush}(\phi
(z))=\partial f_{brush}/\partial \phi $ is the total exchange
chemical potential at $z$, allowing for both the interaction free
energy and the elastic one. This leads to $\mu _{\infty }(\phi
_{+})/kT-Bz_{+}^{2}=\mu _{\infty }(\phi _{-})/kT-Bz_{-}^{2}$.

\begin{figure}[h]
\begin{center}
\epsfig{file=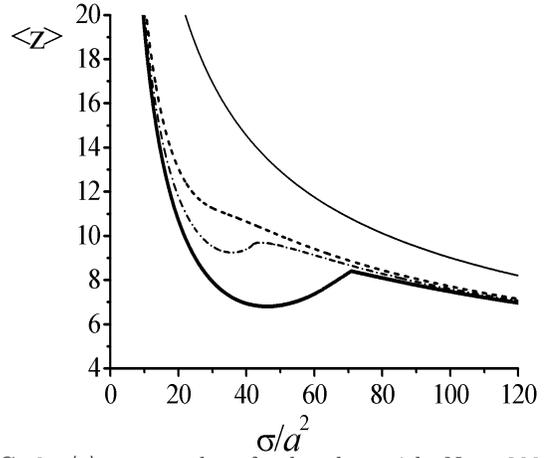, width=7cm, height=6cm} \caption{$\langle
z\rangle $ {\em vs.} $\sigma $ plots for brushes with $N=300$
subject to interactions described by $\overline{\chi }(\phi)=0.5$
(thin line), $\overline{\chi }(\phi)=0.5+0.95\phi^{2}$(dashes),
$\overline{\chi }(\phi)=0.5+1.00\phi^{2}$(dash-dots),
$\overline{\chi} (\phi)=0.5+1.05\phi^{2}$(thick line).}
\end{center}
\end{figure}

At the phase boundary $z_{+}=z_{-}=R_{i}$ and thus $\mu _{\infty
}(\phi _{+})=\mu _{\infty }(\phi _{-})$; (ii) $\pi _{brush}(\phi
_{+})=\pi _{brush}(\phi _{-})$ where $\pi _{brush}=\phi ^{2}\left[
\partial (f_{brush}/\phi)/\partial \phi \right] $ is the local
osmotic pressure. Within the Pincus model, this condition clearly
reduces to $\pi _{\infty }(\phi _{+})=\pi _{\infty }(\phi_{-})$.
Thus, the two coexisting phases are characterized by the monomer
volume fractions, $\phi _{+}$ and $\phi _{-}$, of the bulk phases
in the $ N\rightarrow \infty $ limit. However, because of
(\ref{Nprofle}) the coexistence occurs at a single altitude
$z=R_{i}$ specified by
\begin{equation}
\Delta \mu _{\infty }(\phi _{\pm })/kT=B(R^{2}-z_{i}^{2})
\label{zi}
\end{equation}
thus leading to a discontinuity in $\phi (z)$. Clearly, the
critical point for this transition, as determined by $\partial
^{2}\gamma /\partial \phi ^{2}=\partial ^{3}\gamma /\partial \phi
^{3}=0$, is identical to that of the bulk transition as given by
$\partial ^{2}f_{\infty }(\phi)/\partial \phi ^{2}=0$ and
$\partial ^{3}f_{\infty }(\phi)/\partial \phi ^{3}=0$.

Two approaches allow to explicitly calculate $\phi (z)$ when
$\sigma >\sigma _{c}$. In one, the concentration profile is
obtained from (\ref{Fprofile}) for the intervals $\phi _{0}\geq
\phi \geq \phi _{+}$ and $\phi _{-}\geq \phi \geq 0$. To follow
this route it is necessary to first obtain the bulk binodal by
utilizing, for example, the Maxwell equal-area construction on the
van der Waals loop traced by $\mu _{\infty }(\phi)$.\cite{vdW} The
second approach utilizes a Maxwell construction on the van der
Waals loops occurring in the plots of $\phi (z)$ vs.
$z^{2}$.\cite{Wagner} It is equivalent to the first route because
$z^{2}$ is related to $\mu_{\infty }(\phi)$ via (\ref{Nprofle}).

A vertical phase separation as discussed above becomes possible
once $\phi _{0}$ exceeds $\phi _{-}$. To estimate the threshold
grafting density necessary, $\sigma _{c}$ we assume that for
$\sigma <\sigma _{c}$ the brush thickness retains the scaling
behavior of a single phase brush as obtained from the Alexander
model and the SCF theory. For a Gaussian brush $R/a\approx N\left(
a^{2}/\sigma \right)^{1/2}$ while a brush exhibiting self
avoidance obeys $R/a\approx N\left( a^{2}/\sigma \right) ^{1/3}$.
Since $\Delta \mu_{\infty }(\phi _{0})/kT=BR^{2}$ this leads to
$\sigma _{c}\approx \left[ \Delta \mu _{\infty }(\phi
_{-})/kT\right] ^{-3/2}$ in the self-avoidance case and to $\sigma
_{c}\approx \left[ \Delta \mu _{\infty }(\phi _{-})/kT\right]
^{-1}$ for the Gaussian one. This estimates can serve as
guidelines when $ \Delta \mu _{\infty }(\phi_{0})$ is known, that
is when $\overline{\chi } (\phi)$ is available. When this is not
the case one may roughly estimate $ \sigma _{c}$ by further
assuming that the brush is sufficiently dilute to ensure $\Delta
\mu _{\infty }(\phi)\sim \phi $ thus leading to
\begin{equation}
\sigma _{c}\approx \left\{
\begin{array}{l}
\phi _{-}^{-3/2}\qquad \text{ self-avoidance} \\
\phi _{-}^{-1}\qquad \text{ Gaussian}
\end{array}
\right. \label{sigmac}
\end{equation}
This last form is of interest because it permits a crude estimate
of $\sigma _{c}$ on the basis of the phase diagram even when
$\overline{\chi }(\phi)$ is unknown.

The predicted discontinuity in $\phi (z)$ may prove difficult to
resolve experimentally. A more robust signature of the vertical
phase separation concerns the dependence of the brush thickness on
the grafting density. The vertical phase separation results in
non-monotonous dependence of the moments of the profile on $\sigma
$. For brevity we consider only the first moment of $\phi (z)$
\begin{equation}
\langle z\rangle =\frac{\int_{0}^{R}z\phi (z)dz}{\int_{0}^{R}\phi
(z)dz}= \frac{\sigma }{Na^{3}}\int_{0}^{R}z\phi (z)dz
\label{firstmom}
\end{equation}
and illustrate its $\sigma $ dependence for the hypothetical case
considered in section II, that is of $\overline{\chi
}(\phi)=1/2+\overline{\chi } _{2}\phi ^{2}$. In this case the
critical point is specified by $\overline{ \chi }_{2c}=1$ and
$\phi _{c}=1/2$ so that phase separation occurs when $
\overline{\chi }_{2}\geqslant \overline{\chi }_{2c}$ and $\phi
>\phi _{c}$.

Representative $\phi (z)$ curves for $\overline{\chi }(\phi)$:
$\overline{ \chi }(\phi)=1/2$, $\overline{\chi
}(\phi)=1/2+0.95\phi ^{2}$, $\overline{ \chi }(\phi)=1/2+\phi
^{2}$ and $\overline{\chi }(\phi)=1/2+1.05\phi ^{2}$ are depicted
in Figure 5. The corresponding $\langle z\rangle $ {\em vs.} $
\sigma $ plots are shown in Figure 6. The $\langle z\rangle $ {\em
vs.} $ \sigma $ curves exhibit a pronounced minimum when the brush
undergoes a vertical phase separation while in the single phase
state $\langle z\rangle $ increases monotonically with $\sigma $.
The behavior of $\langle z\rangle $ {\em vs.} $\sigma $ plot
reflects the repartition of the monomers between the two phases.

When $\sigma >\sigma _{c}$ an inner dense phase, with the
associated discontinuity in $\phi (z)$, appears. As a result, the
corresponding increase in the average density within the brush is
mostly spent on the formation of this inner phase. The
contribution of this denser phase to $\langle z\rangle $ is
however weighted by $z$ thus causing the initial decrease in
$\langle z\rangle$.

\section{An Illustrative Example-The case of PNIPAM}

As we have seen, qualitatively novel scenarios for the collapse of
isolated coils and for the structure of polymer brushes occur when
$\overline{\chi } (\phi)$ increases with $\phi $. This behavior is
apparently realized by Poly(N-isopropylacrylamide) (PNIPAM) in
water, a system exhibiting a lower critical solution temperature
(LCST) around $30^{0}C$. Three items concerning this system are of
interest for our discussion.

\begin{figure}[h]
\begin{center}
\epsfig{file=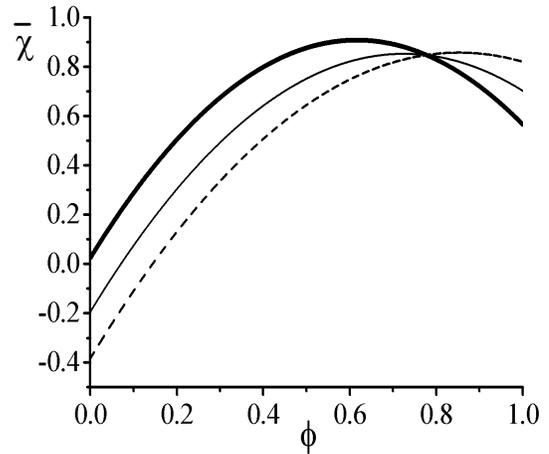, width=7cm, height=6cm} \caption{A plot of
$\overline{\chi }(\phi)$ {\em vs}. $\phi $ for PNIPAM as described
by the $\chi _{eff}(\phi)$ of Afroze {\em et al.}\cite{A,foot2}
for $T=26^{0}C$ (dash), $T=28^{0}C$ (thin line) and $T=30^{0}C$
(thick line).}
\end{center}
\end{figure}

First, is an early study by Zhu and Napper\cite{Napper} of the
collapse of PNIPAM brushes grafted to latex particles immersed in
water. This revealed a collapse involving two stages. An ``early
collapse'', took place below $30^{0}C$, at better than
``$\theta$-conditions'', and did not result in flocculation of the
neutral particles. Upon raising the temperature to worse than
``$\theta$-conditions'' the collapse induced flocculation. This
indicates that the colloidal stabilization imparted by the PNIPAM
brushes survives the early collapse. It lead to the interpretation
of the effect in terms of a vertical phase separation within the
brush due to a second type of phase separation as predicted by the
$n$-cluster model. The second item is the experimental study, by
Wu and his group, of the collapse behavior of isolated PNIPAM
chains.\cite{Wu} This study concerned dilute aqueous solutions of
high molecular weight PNIPAM heated to above the ``$\theta
$-temperature''. It provided detailed $R$ {\em vs.} $T$ plots
characterizing the collapse of individual chains. This study was
made possible by the apparent decoupling of the collapse and bulk
phase separation in the case of PNIPAM. For the present
discussion, the conclusion of interest is that the collapse was
steeper than expected on the basis of a driving force due to
simple binary attractions as modeled by $-v\phi ^{2}$ with
$v=1-2\chi (T)$. The last item concerns $\chi _{eff}(T,\phi)$ and
the phase diagram of PNIPAM. An early study of the phase behavior
of PNIPAM in water, by Heskins and Guillet,\cite{HG} identified an
LCST at $\phi _{c}\simeq 0.16$ and $T_{c}\simeq 31.0^{0}C$. A
recent investigation, by Afroze {\em et al.}\cite{A} led to
different results: (i) While the LCST of PNIPAM depends on $N$ the
LCST occurs around $T_{c}\simeq 27-28^{0}C$ and $\phi _{c}\simeq
0.43$ (ii) In the limit of $\phi \rightarrow 0$, the phase
separation occurs, depending on $N$, between $30^{0}C$ and
$34^{0}C$. Afroze {\em et al }than proceeded to obtain $\chi
_{eff}(T,\phi)$ assuming that it is well described by $\chi
_{eff}(T,\phi)=\sum_{i=0}^{2}\widehat{\chi }_{i}(T)\phi ^{i}$ with
$\widehat{\chi }_{i}(T) $ of the form $\widehat{\chi
}_{i}(T)=A_{i}+B_{i}T$.

\begin{figure}[h]
\begin{center}
\epsfig{file=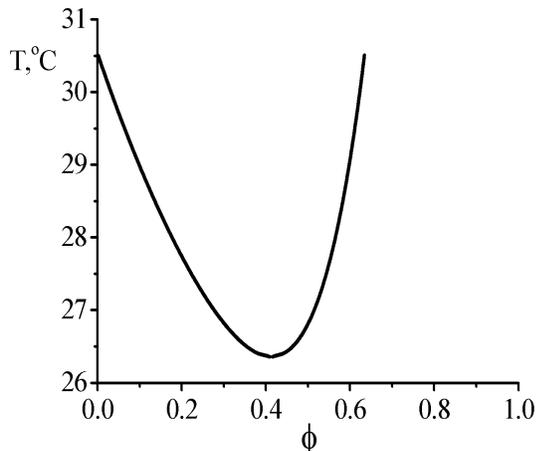, width=7cm, height=6cm} \caption{The phase
diagram of PNIPAM in the limit of $N\rightarrow \infty $ as
obtained from the $\chi _{eff}(\phi)$ of Afroze {\em et
al}.\cite{A,foot2}}
\end{center}
\end{figure}

The six phenomenological parameters were fixed by comparing the
theoretical and experimental phase diagrams.\cite{foot2} In the
following we utilize the results of Afroze {\em et al }because
they are consistent with the results of Zhu and Napper in that
they enable a vertical phase separation within a PNIPAM brush
below $30^{0}C$.

The items listed above suggest that PNIPAM indeed exhibits the
collapse behavior expected when $\overline{\chi }(\phi)$ increases
with $\phi $, thus: (i) Zhu and Napper provided evidence for the
occurrence of a vertical phase separation within a PNIPAM brush;
(ii) Wu and collaborators demonstrated that the collapse of single
PNIPAM is observable and that it is steeper than expected thus
suggesting $\overline{\chi }(\phi)$ involvement; (iii) The phase
diagram measured by Afroze {\em et al} permits the interpretation
of Zhu and Napper concerning PNIPAM brushes. Furthermore, as we
shall see, their $\chi _{eff}(T,\phi)$ yields a $\overline{\chi
}(\phi)$ increasing with $\phi $, thus lending further support to
(i) and (ii).

The $\overline{\chi }(\phi)$ {\em vs.} $\phi $ plots at
$T=26^{0}C$, $T=28^{0}C$ and at $T=30^{0}C$, as obtained on the
basis of $\chi _{eff}(T,\phi)$ proposed by Afroze {\em et
al.}\cite{foot2} (Figure 7) show that $\overline{\chi }(\phi)$
increases with $\phi $. The phase diagram for $N\rightarrow \infty
$, (Figure 8) as calculated using this $\chi _{eff}(T,\phi)$,
exhibits a second type of phase separation, as stated by Afroze
{\em et al.} The concentration profiles of a PNIPAM brush, $\phi
(z)$ {\em vs.} $z$, thus obtained confirm that a vertical phase
separation is indeed expected within the brush (Figure 9). The
above plots suggest that PNIPAM in water is indeed a system which
exhibits the novel signatures associated with $\overline{\chi
}(\phi)$ that increases with $\phi $ and with a second type of
phase separation. At the same time, it is important to stress that
the performance of $\chi _{eff}(T,\phi)$ proposed by Afroze {\em
et al.} is not faultless. Using this $\chi _{eff}(T,\phi)$ enabled
Afroze {\em et al }to reproduce satisfactorily only one of the
four phase diagrams they studied. With this in mind, the plots in
Figures 7--9 should be considered as preliminary.

\begin{figure}[h]
\begin{center}
\epsfig{file=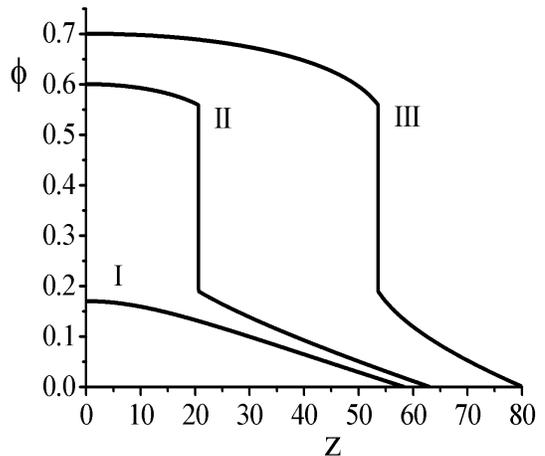, width=7cm, height=6cm} \caption{$\phi $
{\em vs}. $z$ plots for PNIPAM brushes for $N=300$ and $T=28^{0}C$
as obtained from the $\chi _{eff}(\phi)$ of Afroze {\em et
al.}\cite{A,foot2} Curves I, II and III correspond respectively to
$\sigma /a^{2}=53$ ($\sigma /R_{F}^{2}=0.056$), $\sigma /a^{2}=19$
($\sigma /R_{F}^{2}=0.020$), $\sigma /a^{2}=8$ ($\sigma
/R_{F}^{2}=0.009$).}
\end{center}
\end{figure}

Hopefully, better results can be obtained when direct measurements
of $\overline{\chi }(T,\phi)$ for PNIPAM in water will become
available.

\section{Discussion}

The necessity of introducing $\chi _{eff}(T,\phi)$ to replace
$\chi =\chi (T)$ signals a failure of the Flory free energy. This
failure is traceable to deficiencies in the Flory-Huggins lattice
model and shortcomings of the approximations invoked in obtaining
the Flory free energy. Thus, the Flory-Huggins lattice model
assumes that all monomers are identical in size and shape to the
solvent molecules. Furthermore, it supposed that all monomers
exist in a single state. In turn, the Flory approximation fails to
distinguish between monomer-monomer contacts due to intrachain
contacts and those due to interchain ones. The resulting free
energy is insensitive to the monomer sequence of heteropolymers.
It only allows for pairwise attractive interactions and not for
higher order attractions involving micelle like clusters of
monomers. These deficiencies motivated a number of theoretical
refinements of the Flory-Huggins lattice
model\cite{n-Cluster,Freed,Painter,tanaka,BBP,K} and of its
solution. These refinements are achieved at the price of
introducing additional parameters. In certain cases, it is
conceivable that a full description of a system may require a
combination of a number of these refined treatments. For example,
in modelling neutral water soluble polymers it may be necessary to
allow for monomer size and shape, the existence of interconverting
monomeric states and for the interplay of intrachain and
interchain contacts. The number of parameters involved with such a
complete description is even higher. With this in mind, it is of
interest to utilize a complementary approach involving the Flory
free energy and the measured $\overline{\chi }(T,\phi)$, as
obtained from the colligative properties of the polymer solutions.
In this paper we focused on the relationship between
$\overline{\chi }(T,\phi)$, a macroscopic property, and the
microscopic swelling and collapse behavior of coils and brushes.
As we have seen, this approach yields specific predictions
concerning: (i) the cross-over between ideal chain and self
avoidance statistics, $g_{B}$; (ii) the concentration profile,
$\phi (z)$, of a polymer brush. It also suggests the possibility
of a first order collapse transition for flexile chains when
$\overline{\chi }(T,\phi)$ increases with $\phi $. This route has
the merit of relating independent measurements and helps clarify
the significance of $\overline{\chi }(T,\phi) $ and $\chi
_{eff}(T,\phi)$ as measures of solvent quality. Clearly, this
phenomenological approach does not yield insights concerning the
molecular origins of $\overline{\chi }(T,\phi)$ and $\chi
_{eff}(T,\phi)$. Consequently, it does not identify molecular
design parameters allowing to tune $\overline{\chi }(T,\phi)$ and
$\chi _{eff}(T,\phi)$. The applicability of this method is also
limited by the paucity of systematic tabulations of
$\overline{\chi }(T,\phi)$. To clarify the advantages and
disadvantages of this method it is helpful to consider the
vertical phase separation within a brush as considered in sections
V and VI. In order to design an experiment yielding direct
evidence for this transition, it is necessary to identify a
polymeric system exhibiting such effect. Another prerequisite is a
clear idea of the range of grafting densities, molecular weights
and temperatures involved. The effect was originally predicted
within the framework of the $n$-cluster model proposed for PEO.
This model was subsequently invoked in the interpretation of the
indirect evidence for this transition in brushes of PNIPAM.
However, utilizing the $n$-cluster model as a guide for further
studies is hampered by two problems: (i) the applicability of the
$n$-cluster model to this system vis-\`{a}-vis the alternative
model of Matsuyama and Tanaka\cite{tanaka2} is not clear; (ii) the
$n$-cluster model, like the competing approach of Matsuyama and
Tanaka, invokes parameters that are currently unknown. On the
other hand, as we have seen, the necessary information can be
obtained directly from experimentally determined $\chi
_{eff}(T,\phi)$ (or equivalently, $\overline{\chi }(T,\phi)$). The
implementation of this last approach is clearly limited by the
accuracy of the reported $\overline{\chi}(T,\phi)$ and
$\chi_{eff}(T,\phi)$. This last difficulty, concerning the
uncertainties $\overline{\chi }(T,\phi)$ and the phase diagram of
PNIPAM, hampers however also the microscopic models that require
such data in order to determine the parameters of the theory.

\end{multicols}

\begin{references}
\bibitem{nom} In the literature,\cite{wolf} $\chi _{eff}$ is sometimes
denoted by $g$ while $\chi $ denotes $\overline{\chi }=\chi _{eff}-(1-\phi
)\partial \chi _{ef}/\partial \phi $. Our notation aims to avoid confusion
with the customary usage of $\chi $.
\bibitem{footX} For brevity we will often replace $\chi _{eff}(T,\phi)$ and
$\overline{\chi }(T,\phi)$ by $\chi _{eff}(\phi)$ and
$\overline{\chi} (\phi)$.
\bibitem{chiefff} $\chi_{eff}$ is obtained by integrating $\overline{\chi}
(\phi)=-\frac{\partial}{\partial\phi}\left[\chi_{eff}(\phi)(1-\phi)\right]$.
\bibitem{muexp} Upon expressing $\chi _{eff}(\phi)$ in terms of $\overline{\chi}(\phi)$ we obtain $\mu _{\infty }(\phi)=-\ln (1-\phi
)-1+\chi _{eff}(0)-\int_{0}^{\phi }\overline{\chi }(\phi
^{^{\prime }})d\phi ^{^{\prime }}-\phi \overline{\chi }(\phi)$.
\bibitem{vdW} The equality $\pi _{\infty }=\phi \mu _{\infty }-f_{\infty}$
leads to $\partial \pi _{\infty }/\partial \phi =\phi \partial \mu
_{\infty }/\partial \phi $ and thus to $\pi _{\infty }(\phi
_{+})-\pi _{\infty }(\phi _{-})=\int_{\phi _{-}}^{\phi _{+}}\left(
\phi \partial \mu _{\infty }/\partial \phi \right) d\phi $. Since
the binodal is determined by $\pi _{\infty }(\phi _{+})=\pi
_{\infty }(\phi _{-})$ and $\mu _{\infty }(\phi _{+})=\mu_{\infty
}(\phi _{-})$ the equal area construction is specified by $\mu
_{\infty }(\phi _{\pm })(\phi _{+}-\phi _{-})=\int_{\phi
_{-}}^{\phi _{+}}\mu _{\infty }(\phi)d\phi$.
\end{references}
\end{document}